\documentclass[conference]{IEEEtran}
\IEEEoverridecommandlockouts

\usepackage{cite}
\usepackage[table]{xcolor}     
\usepackage{amsmath,amssymb}
\usepackage{graphicx}
\usepackage{textcomp}
\usepackage{booktabs}
\usepackage{caption}
\usepackage{subcaption}
\usepackage{url}
\usepackage[ruled,linesnumbered]{algorithm2e}  
\usepackage{mathptmx}
\usepackage[colorlinks=true, allcolors=blue]{hyperref}

\def\BibTeX{{\rm B\kern-.05em{\sc i\kern-.025em b}\kern-.08em
    T\kern-.1667em\lower.7ex\hbox{E}\kern-.125emX}}

\begin{document}

\title{Priority-Aware Deficit Round Robin for AXI4 Bandwidth Fairness in ASAP-7nm Process}

\author{
\IEEEauthorblockN{Tasmin Khan}
\IEEEauthorblockA{\textit{Dept. of EEE} \\
\textit{Bangladesh Univ. of Eng. \& Tech.}\\
tasminkhanelec@eee.buet.ac.bd}
\and
\IEEEauthorblockN{Tishna Sabrina}
\IEEEauthorblockA{\textit{Dept. of EEE} \\
\textit{Univ. of Asia Pacific}\\
tishna@uap-bd.edu}
\and
\IEEEauthorblockN{Ch. Md. Rakin Haider}
\IEEEauthorblockA{\textit{Dept. of CSE} \\
\textit{Bangladesh Univ. of Eng. \& Tech.}\\
rakinhaider@cse.buet.ac.bd}
}
\maketitle


\begin{abstract}
Modern systems-on-chip place real-time clients, processor cores and bulk
data engines on the same AXI4 interconnect, where they contend for
shared slaves. AXI4 fixes the ordering rules and carries a
4-bit quality of service (QoS) field, but leaves the service policy to
the implementer, and that policy determines how long each master waits
and how much bandwidth it receives once a slave saturates. We propose a
QoS-aware deficit round robin (DRR) architecture that charges each
grant in beats, and benchmark it against two architectures
built behind the same port list: round robin, and QoS priority with
anti-starvation aging. Charging in beats recovers bandwidth fairness
while leaving every burst whole, so the address phase is issued once and
locality for memory access is preserved. All three are driven by the
same deterministic traffic generator and taken from RTL to GDSII on the
ASAP7 7-nm predictive PDK. Under a 1:16 burst-length mismatch, the round robin and QoS
architectures both hold the beat ratio between the two competing
masters at 1:16, while our design brings it to 1:1 at full load. Mean read latency for the
priority master falls from 95 cycles under round robin to 67 with QoS
and 37 with QoS-aware DRR. At 500 MHz this
serves the priority master in 74 ns against 190 ns. The
measurements give a per-policy price and a traffic-based guide for
choosing among the three.
\end{abstract}


\section{Introduction}\label{sec:intro}
Heterogeneous systems-on-chip (SoCs) connect processor cores,
direct-memory-access (DMA) engines, and domain-specific accelerators to
shared memories and peripherals through burst-based on-chip
interconnects, and AMBA AXI4 is the prevailing protocol for these
fabrics~\cite{arm_axi,kurth2022}. The AXI4 specification defines the
channels, handshakes, and ordering rules of a transaction. However, it leaves the
arbitration among contending masters to the
implementer~\cite{restuccia2019fair}. Once a shared slave saturates, that
arbitration policy decides how long each master waits and how much
bandwidth each master receives.
 
Two traffic situations make this choice consequential. In the first, a
latency-critical master \cite{arm_qos400,benz2025realm} with little data shares a slave with masters
that move bulk data. A processor stalled on a cache miss depends
directly on transaction latency, while DMA controllers and graphics
processors issue traffic to the same memory
controller~\cite{arm_qos400}. Mixed-criticality automotive SoCs show the
same pattern, with short cache-line-sized transactions from cores
competing against long bursts from accelerators~\cite{benz2025realm}.
AXI4 attaches a 4-bit QoS identifier to each address request for use as
a priority indicator~\cite{arm_axi}, and this can be used as an arbitration priority~\cite{arm_qos400}. A priority arbiter admits the critical request first. It cannot interrupt a long burst
already in service, and it needs an aging mechanism to keep
low-priority masters from starving~\cite{silberschatz2018}.
 
In the second situation, masters of equal priority issue bursts of very
different lengths. Round-robin arbitration, the common default in AXI
interconnects, is fair in the number of grants, so a master receives
bandwidth in proportion to its burst length. On FPGA SoCs, seven
interferers issuing 256-beat bursts reduce a victim accelerator to under
1\% of the bus bandwidth~\cite{restuccia2019fair}, and the same effect
appears in bus arbitration for real-time multicore
processors~\cite{slijepcevic2017}. A router in a packet-switched network faces the same contention where multiplexing variable-length packets from many flows onto one shared link. Fair queueing solved it by serving each flow in proportion to its data rather than its packet count~\cite{demers1989}, and deficit round-robin (DRR) is the low-cost approximation where each flow holds a per-round quantum and a running deficit~\cite{shreedhar1996}. An AXI4 burst is the on-chip counterpart of a packet, holding the slave for as many beats as its length. DRR alone meters bandwidth but carries no notion of urgency. So we combine this with QoS, so that priority and bandwidth share are set by separate controls. In our interconnect design, a per-transaction QoS value orders who is served first, and a per-master weight sets each master's beat quantum.

Existing AXI4 work attacks these problems in multiple ways. Some reshape the traffic before it reaches the arbiter such as the burst equalizer splits long bursts so a grant-counting arbiter divides bandwidth evenly~\cite{restuccia2019fair}. AXI-REALM fragments bursts at the ingress under a per-manager budget~\cite{benz2024realm,benz2025realm}. Some throttle the source using credit and reservation schemes to cap each master's share over a time window~\cite{slijepcevic2017,pagani2019}. Some leave transaction admission untouched and reorder them once buffered~\cite{jiang2023axiicrt}. Splitting a burst into $n$ sub-requests multiplies address-channel occupancy by $n$ and requires the returning
read data to be reassembled~\cite{restuccia2019fair}. Addtionally it presents the
downstream memory controller with $n$ short accesses in place of one,
where row-buffer locality sets the achievable efficiency. So, we design the arbiter to meter bandwidth by beats while still granting by priority, without reshaping any transaction or an extra regulator upstream. The read queue then serves buffered bursts by that same priority.

\begin{figure*}[t]
\centering
\includegraphics[width=\linewidth]{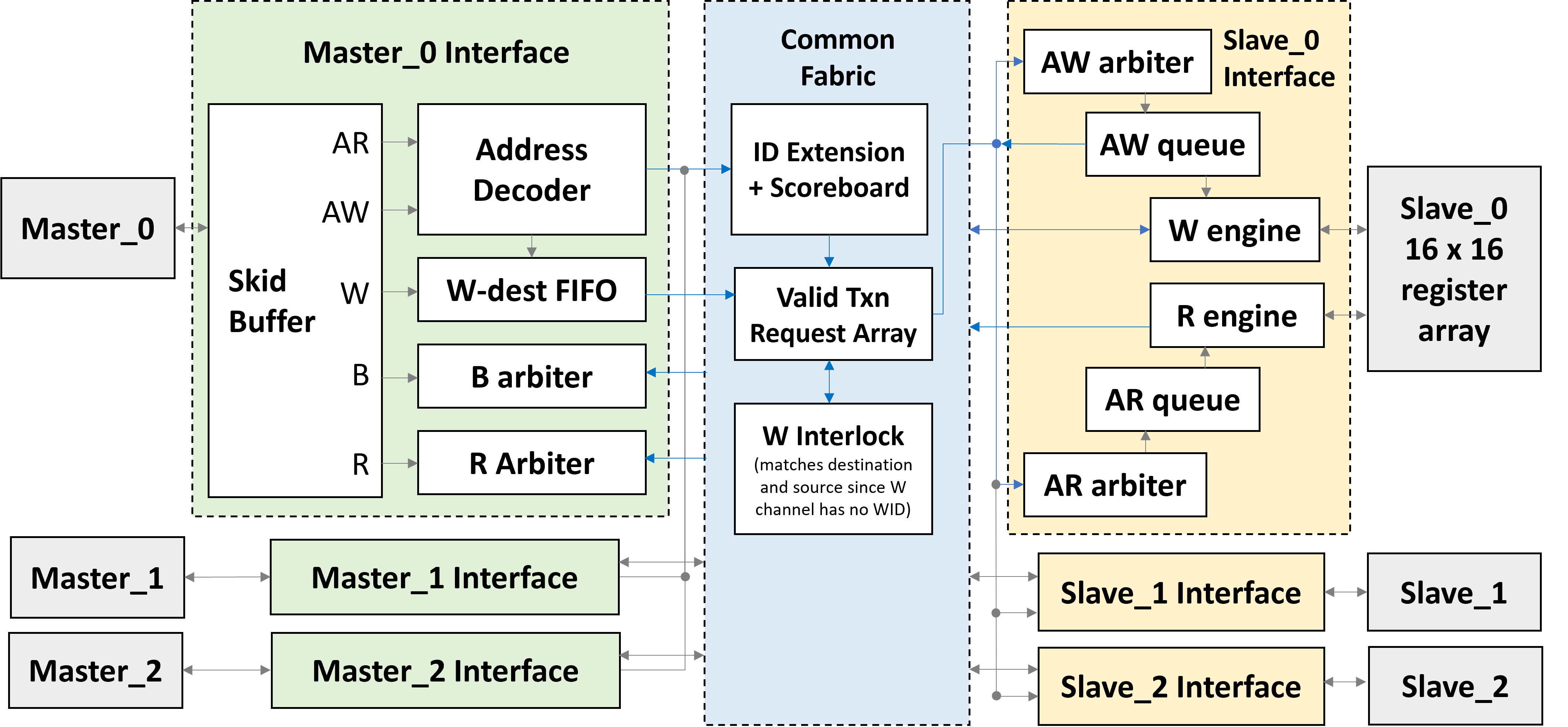}
\caption{Microarchitecture of the $3\times3$ AXI4 crossbar. Same master and slave
interfaces are replicated per port.}
\label{fig:arch}
\end{figure*}

We build three service scheduling
architectures for a $3\times3$ AXI4 crossbar and take each through RTL-to-GDSII on ASAP7 7-nm predictive PDK~\cite{clark2016asap7}. Our contributions
are:
\begin{itemize}
\item A QoS-aware DRR arbiter for AXI4 that keeps priority and
bandwidth share as separate controls. Priority travels per transaction
in AxQOS, and share is a per-master weight held in configuration. A
master can therefore be high-priority and low-bandwidth, and
equal-priority masters can receive weighted shares. The arbiter charges
each grant in beats and needs no aging counters.
\item A head-to-head physical implementation of RR, QoS with aging, and
QoS+DRR from one RTL base under identical constraints, reporting cell
area, power, critical-path delay, and wire length for each policy at an advanced nod.
\item A traffic-based decision guideline, backed by RTL simulation under
identical stimulus, that assigns each policy to the traffic class it
serves best.
\end{itemize}

\section{Crossbar Architecture}\label{sec:arch}

AXI4 splits a transaction across five channels; addresses (AW, AR), write data (W), and responses (B, R). Address and response channels carry an ID; W does not, so write data is bound to its address only by order \cite{arm_axi}. Fig.~\ref{fig:arch} shows the interconnect architecture as designed. It has a per-master interface, a common fabric for shared bookkeeping, and a per-slave interface, with the three policies of Section~\ref{sec:sch} differing inside the arbiters and the AR queue.

\subsection{Master Interface}

Each of the five channels is buffered by a skid buffer. The address decoder uses buffered AW/AR address to issue a target-slave
select, or an error code when the address is unmapped or the burst would cross
a slave boundary. The fabric widens every ID with a
tag identifying the originating master. This tag is later used for
the return path. The B and R arbiters allow the beats to be claimed by the master whose index matches its tag. So, different masters may accept responses from different slaves in the same cycle.

AXI4 requires a master to issue W data in AW order, so a W-destination FIFO is filled in AW-grant order and drained per W burst. When a master's AW is granted to a slave, that slave's index is pushed; the head names where this master's \emph{next} W burst is steered.

\subsection{Common Fabric}

The ID extension block extends the ID with a tag and drives a scoreboard that permits one outstanding transaction per ID. AXI4 allows responses out of order across different IDs but requires same-ID ordering. One transaction per ID guarantees no two in-flight requests share an ID, so any reordering the fabric performs is protocol-legal. The valid-transaction request array is the masters$\times$slaves eligibility grid presented to the arbiters. A master is a candidate at a slave when its decoded target matches, its tagged ID is free, and, for writes, its W-destination FIFO has room. Requests blocked upstream are withheld from arbitration, so the anti-starvation guard of Section~\ref{sec:sch} do not mistake a scoreboard stall for starvation.

Two things fix the W routing: the master's W-destination FIFO says which slave its data goes to, and the slave's AW queue head says which master that slave is currently accepting. The W interlock matches these to forward one master's W beats at a time, till the end of the burst.

\subsection{Slave Interface}

Each address channel has a per-slave arbiter that selects one master from the request array and drives the slave port. Accepted addresses wait in a queue until the corresponding engine can service them. The read and write channels are independent, so a master reading back a value it just wrote must wait for the B response before issuing the read. The AW queue is a strict FIFO in all three. Reordering it would make the interlock's source tag name a different burst than the one whose data is arriving, and since W beats carry no ID the mismatch would commit the wrong master's data with nothing to catch it. The AR queue is a FIFO in the round-robin architecture and reorderable in the other two, where an entry is selected by policy rather than by arrival order.

The write engine consumes W beats from the interlock, commits them to the register array, and emits one B response at the end of the burst. The read engine services one AR entry at a time, streaming its beats to the R channel until the final beat staying locked to a burst. Service scheduling policies therefore act at burst boundaries, choosing which transaction is served next, and never preempts one in flight. Each slave backs a $16\times16$-bit register array, kept small to focus on measuring interconnect fabric cost rather than memory-macro area. An address that the decoder marks as unmapped or boundary-crossing is routed to an error slave, which drains the burst inputs and returns a DECERR response.

\section{Traffic Service Policy}\label{sec:sch}

In the fabric of Section~\ref{sec:arch}, several masters can address the same slave
and one master can receive responses from multiple slaves at the same
time. A policy therefore acts at two points. At \emph{admission}, the
per-slave arbiter chooses which contending master enters the slave's
queue. At \emph{service}, the read engine chooses which queued request to
run next. That choice sets the
latency each master observes and how the interconnect's bandwidth is divided.
It takes effect only under contention: below saturation knee requests rarely
queue and every policy behaves alike. A
latency-sensitive master must be served promptly but consumes little
bandwidth, while a bandwidth-heavy master issues long bursts and
tolerates delay.

We implement three scheduling architectures and compare them under
identical traffic. Each makes the
admission and read-reorder decisions differently in three hooks:
\texttt{Eligible}, which decides who may compete; \texttt{Pick}, which
chooses the winner among them; and \texttt{OnGrant}, the state update
after a grant is accepted (Algorithm~\ref{alg:hooks}). On the
read path the slave's reorder pool applies the same \texttt{Pick} rule
to choose the next queued burst, while the write path stays strictly in address issue
order in all three. 

\textbf{Round-Robin (RR):}
Every requester is eligible, and \texttt{Pick} takes the first requester
encountered scanning from the round-robin pointer, which advances past
the winner on each accepted grant. \texttt{OnGrant} keeps no state, and
the read queue is served in arrival order. Grants are fair per
request count, so a master issuing longer bursts receives more beats.

\textbf{Quality of Service (QoS) Priority:}
Every requester is eligible. \texttt{Pick} takes the highest priority, which equals a request's \textsc{arqos}/\textsc{awqos} value
unless the master has waited \textsc{starve\_limit} rounds, then starved master's priority is 
raised to the maximum. \texttt{OnGrant} clears the winner's
age and increments the age of every master still waiting. The read pool
selects the next burst by the same effective-priority rule. The aging term bounds a low-priority master's wait, which higher-priority requests would otherwise extend indefinitely \cite{silberschatz2018}.

\begin{algorithm}[t]
\caption{Per-slave address arbitration.}
\label{alg:arb}
\DontPrintSemicolon
\SetKwFunction{Elig}{Eligible}
\SetKwFunction{Pick}{Pick}
\SetKwFunction{OnGrant}{OnGrant}
\KwIn{per-master valid $v_m$, payload $\{id,addr,len,qos\}_m$, ready $r$}
\KwOut{winner $g$}
\eIf{\textup{hold valid} \textbf{and} $v_{\mathrm{hold}}$}{
  $g \leftarrow \mathrm{hold\_idx}$ \tcp*{offered payload cannot change}
}{
  $E \leftarrow \{\, m : v_m \wedge \Elig{m} \,\}$\;
  $g \leftarrow \Pick{E}$\;
}
drive payload$_g$ to slave; assert ready to master $g$ only\;
\If{$g$ granted \textbf{and} $r$}{
  $\mathrm{rr\_ptr} \leftarrow (g+1)\bmod N$\;
  \OnGrant{$g$}\;
}
\end{algorithm}

\begin{algorithm}[t]
\caption{Three architectures hooks into Alg.~\ref{alg:arb}.}
\label{alg:hooks}
\DontPrintSemicolon
\LinesNotNumbered
\SetKwProg{Pol}{}{:}{}
\Pol{RR}{
  \texttt{Eligible}$(m)=\textbf{true}$\;
  \texttt{Pick}$(E)=$ first requester from rr\_ptr\;
  \texttt{OnGrant}$(g)=$ none\;
}
\vspace{2pt}
\Pol{QoS (priority with aging)}{
  \texttt{Eligible}$(m)=\textbf{true}$\;
  $\mathit{eff}(m)=\textsc{max}$ if $\mathit{age}(m)\ge\textsc{limit}$,
  else $\mathit{qos}(m)$\;
  \texttt{Pick}$(E)=\arg\max_{m\in E}\mathit{eff}(m)$\;
  \texttt{OnGrant}$(g)$: $\mathit{age}(g)\!\leftarrow\!0$;
  age of others still waiting $\mathrel{+}{=}1$\;
}
\vspace{2pt}
\Pol{QoS+DRR}{
  $\mathit{cost}(m)=\mathit{len}(m)+1$ \tcp*{beats}
  \texttt{Eligible}$(m)=[\,\mathit{deficit}(m)\ge\mathit{cost}(m)\,]$\;
  \texttt{Pick}$(E)=\arg\max_{m\in E}\mathit{qos}(m)$ (rr\_ptr breaks ties)\;
  \texttt{OnGrant}$(g)$: $\mathit{deficit}(g)\mathrel{-}{=}\mathit{cost}(g)$\;
  \textbf{round end} (some $v_m$, none eligible):
  $\mathit{deficit}(m)\mathrel{+}{=}\mathit{quantum}(m)$ per requester\;
}
\end{algorithm}

\textbf{Deficit Round-Robin with QoS (QoS+DRR).}
Under RR a master issuing long bursts wins the same number of grants as
one issuing single-beat requests, so the former takes a larger share of bandwidth. To make
it fair per beat, each master holds a deficit counter and a per-round
credit as its quantum
$\mathit{quantum}(m)=\textsc{q\_min}+\mathit{weight}(m)\times\textsc{step}$, where $\mathit{weight}(m)$ is a per-master constant that sets its share
of bandwidth.
A grant costs $\mathit{cost}(m)=\mathit{len}(m)+1$ beats, and a master is
eligible only while its deficit covers that cost. Among eligible masters
\texttt{Pick} takes the highest priority, a round-robin pointer
breaks ties, and \texttt{OnGrant} subtracts the granted cost from the
winner's deficit. When masters are requesting but none is eligible, the
round ends and every requester is credited its quantum. Charging in beats ties each master's bandwidth share to its quantum, and so to its predefined weight.

The quantum floor is a correctness condition. Every master's quantum
must cover its largest possible burst,
\begin{equation}
Q_{\min}\;\ge\;\text{max burst cost}\;=\;\textsc{len}_{\max}+1,
\label{eq:qfloor}
\end{equation}
otherwise a master whose burst exceeds its quantum can never accumulate enough
deficit to become eligible because the deficit is not carried beyond the round in which credit is granted. Each requester is credited its full quantum
at every round end, which bounds the worst-case wait to one round. The
QoS+DRR architecture therefore carries no age counters.

\section{Methodology}\label{sec:method}

The three architectures are separate RTL trees with identical top-level
ports. Every functional
and physical run uses the same stimulus, constraints, and technology.

\subsection{Traffic Simulation}
Performance is measured on the read path under sustained contention. All
three masters read from Slave~0 so they compete at a single arbiter,
which is where the scheduling policy acts. Master~m0 is the
latency-sensitive client: short bursts (\textsc{arlen}=3), high priority
(\textsc{arqos}=15), and low bandwidth. Masters m1 and m2 are the bulk
pair, with equal weight and no priority (\textsc{arqos}=0). They differ
only in burst length: m1 issues single-beat reads (\textsc{arlen}=0) and
m2 issues 16-beat reads (\textsc{arlen}=15). At equal weight this
$16\times$ burst disparity is the condition that separates per-request
from per-beat fairness. The QoS+DRR architecture uses weights $(w_0,w_1,w_2)=(1,6,6)$ in both
profiles, which gives m1 and m2 equal quanta.

Two profiles are swept. In the latency profile, m0 is held at a fixed
low offer rate while the bulk pair sweeps, and the reported quantity is
each master's average total latency versus offered load. In the fairness
profile, m0 is silent and only m1 and m2 compete, giving the m1:m2 beat
ratio on its own. Offered load is swept from 1\% to 100\% per cycle per master.

Offered load is generated by a per-master Bresenham credit accumulator \cite{bresenham1965}.
Each master accrues credit proportional to its load setting every cycle
and issues a read once its credit reaches one transaction and a free ID
is available; the credit is clamped to bound backpressure banking.
Because injection is deterministic, the offered load is bit-identical
across the three architectures whatever the fabric does downstream.

\subsection{Physical Implementation}
Each RTL tree is taken through an RTL-to-GDSII flow on the ASAP7 7-nm
predictive PDK. The behavioral RTL is synthesized against the technology
\texttt{.lib} with a 2~ns clock period. This target period was chosen since 500\,MHz clock is within the range of high-end microcontroller core \cite{st_stm32h743}. The netlist is floorplanned to 60\% core utilization with master ports on the left side and slave ports on the right, and a power mesh is defined. Timing, Connectivity and DRC are
rechecked after each stage. The implementation setup and constraints are
listed in Table~\ref{tab:constraints} and are held identical across all
three physical runs, so any difference in area, power, timing, or wire
length in section \ref{sec:results} reflects the scheduling logic.

\begin{table}[t]
\caption{Common Physical Implementation Constraints}
\label{tab:constraints}
\centering
\footnotesize
\arrayrulecolor{gray!80}
\setlength{\arrayrulewidth}{0.4pt}
\setlength{\tabcolsep}{5pt}
\renewcommand{\arraystretch}{1.4}
\begin{tabular}{|l|l|l|l|}
\hline
\textbf{Parameter} & \textbf{Value} & \textbf{Parameter} & \textbf{Value}\\
\hline
Technology       & ASAP7            & Delay model      & CCS\\
\hline
Configuration    & 3M, 3S           & Input/output delay & 200\,ps\\
\hline
Clock period     & 2\,ns (500\,MHz) & Core utilization & 60\,\%\\
\hline
Process corner   & Typical (TT)     & Aspect ratio     & 0.6\\
\hline
Supply voltage   & 0.7\,V           & Temperature      & 25\,\textdegree C\\
\hline
\end{tabular}
\arrayrulecolor{black}
\end{table}

\begin{figure*}[t]
\centering
\includegraphics[width=\linewidth]{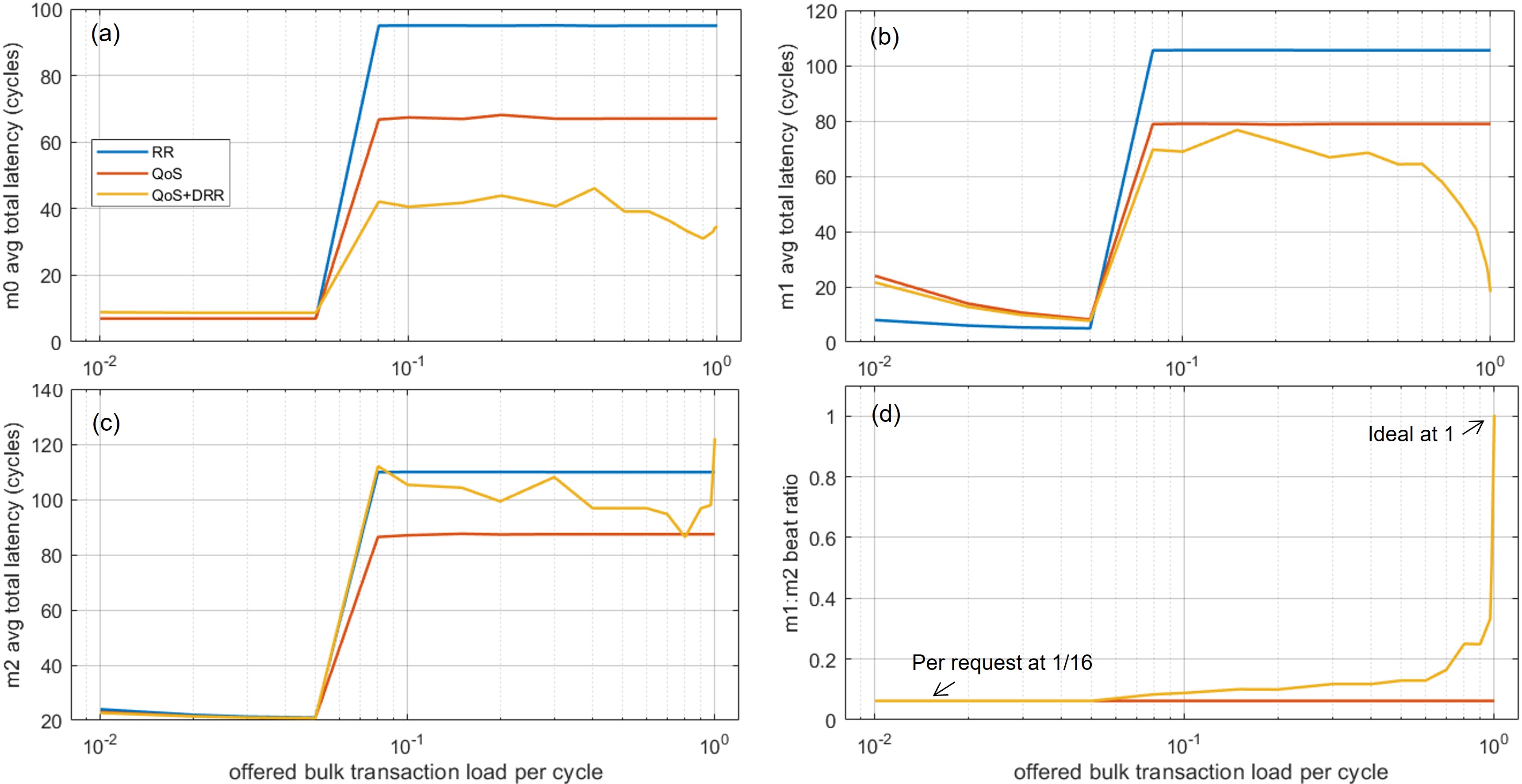}
\caption{Read-path behavior versus the offered load of the bulk transaction (a)--(c) Average total latency of m0, m1 and m2 in the latency profile. (d) m1:m2
delivered-beat ratio in the fairness profile. The RR
curve lies under the QoS curve at $1/16$.}
\label{fig:traffic}
\end{figure*}

\begin{figure}[t]
\centering
\includegraphics[width=\columnwidth]{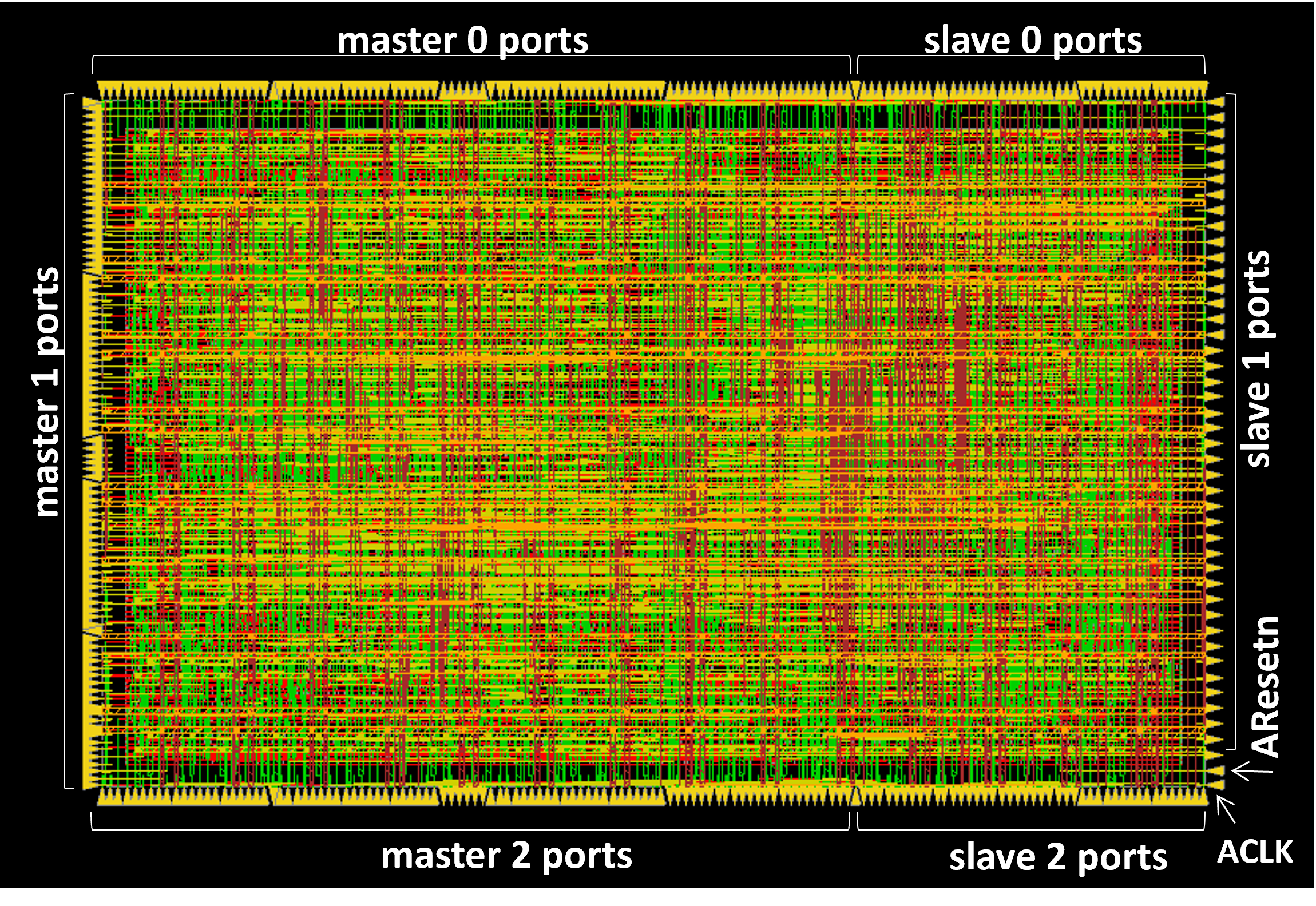}
\caption{Routed layout of the QoS+DRR architecture crossbar. The master ports are on the left and slave ports on the right.}
\label{fig:layout}
\end{figure}

\section{Results and Discussion}\label{sec:results}

\subsection{Latency under Transaction Load}

In the latency profile, the high-priority master m0 issues 4-beat reads at a fixed 1\% rate, once every 100 cycles. The bulk masters m1 and m2 issue reads at equal weight, and their load is swept from 1\% to 100\%. m0 models a latency-critical client such as a sensor interface: it issues rarely but needs each request served quickly. Each point is averaged over a 10{,}000-cycle run. All three architectures saturate between 5\% and 8\% load (Fig.~\ref{fig:traffic}(a)--(c)). At load $L$ the bulk pair offers $17L$ beats per cycle, so demand exceeds the read path's one-beat-per-cycle capacity near $L$= 5.6\%. Below this knee, latency just follows burst length and the three architectures stay within a few cycles of each other. Above it, the read path is full and the policy decides only how the beats and the waiting are shared.

QoS lowers the latency of all three masters above the knee. m0, the only
master carrying a priority value, falls from 95 cycles under RR to 67. However,
m1 and m2 also fall from 105 to 79 and from 110 to 87. This gain comes from faster admission
rather than at service. Serving m0's short reads first turns the slave's AR queue over faster and
every master is admitted sooner. m0 gains at both stages and hence falls the
furthest by 29\% compared to 25\% and 20\% for m1 and m2. 

QoS serves m0 first, but a bulk m2 burst still holds the read engine for 16 beats, and m2 returns to compete immediately. QoS+DRR charges each grant in beats, so a single 16-beat burst empties m2's deficit and locks m2 out until the round ends. For the rest of that round, m0 faces no long-burst rival. The same effect helps m1, whose latency drops from 105 cycles under RR to 47 under QoS+DRR, reaching 18 cycles at full load. And m2 pays for this, rising to 100 cycles, still 8\% below RR. Despite QoS+DRR's longer critical path, this gain survives in real time. At a clock frequency of 500 MHz, m0's saturated latency is 190~ns under RR, 134~ns under QoS, and 74~ns under QoS+DRR.

\begin{figure*}[t]
\centering
\includegraphics[width=\linewidth]{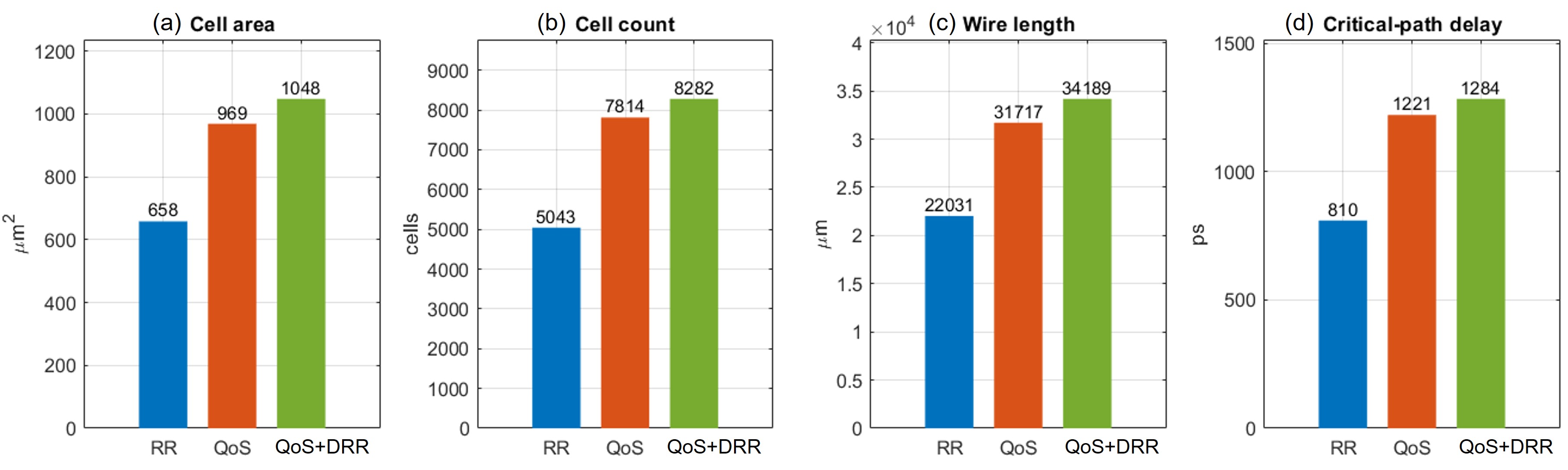}
\caption{Post-route (a) cell area, (b) cell count, (c) total wire length, and (d)
critical-path delay of the three architectures.}
\label{fig:ppa}
\end{figure*}

\begin{figure}[t]
\centering
\includegraphics[width=0.9\columnwidth]{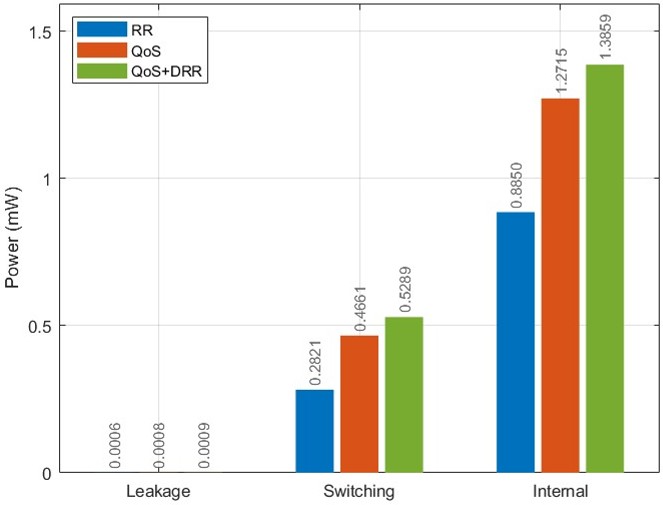}
\caption{Leakage, switching, and internal power of the three
architectures.}
\label{fig:power}
\end{figure}

\subsection{Bandwidth Fairness}\label{sec:fairness}
 
Fig.~\ref{fig:traffic}(d) shows the m1:m2 beat ratio. RR and QoS
hold the ratio at 0.0625 at every load. Both grant per request while m2's bursts
are 16 times longer \cite{restuccia2019fair}. Under QoS+DRR the ratio rises with load, from
0.084 at 8\% to 0.25 at 80\%, 0.33 at 93\%, 0.50 at 98\% and 1.00 at 99--100\%.
At full contention the per-beat charge gives the two equal-weight masters equal
bandwidth. At partial load m1 does not hold a pending request at every round
end, and the quantum is credited only to masters that are requesting, so m1
forfeits part of its share. Total delivered beats stay at 10,000 for all three.

\subsection{Power, Performance and Area (PPA)}
 
Fig.~\ref{fig:layout} shows the routed layout for QoS+DRR architecture. Fig.~\ref{fig:ppa} summarizes the post-route results. Relative to RR, QoS increases cell area by 47\%, instance count by 55\%, wire length by 44\%, total power by 49\% and critical-path delay by 51\%. The increment of QoS+DRR over QoS
is 8.2\% in cell area, 6.0\% in instances, 7.8\% in wire length, 10.2\% in
power and 63~ps in delay. Area per instance stays between 0.124 and
0.131\,$\mu$m$^2$ and wire length per instance between 4.06 and 4.37\,$\mu$m in
all three. The growth in area, wire length and critical path delay, therefore tracks the added
scheduling logic at constant cell size and routing density. Internal power accounts for 72--76\% of the total, switching power for 24--28\% and leakage for less than 0.1\% (Fig.~\ref{fig:power}). 
 
QoS+DRR costs 390\,$\mu$m$^2$ of cell area and 0.75\,mW over RR, and
79\,$\mu$m$^2$ and 0.18\,mW over QoS. For this cost it lowers the saturated
latency of the latency-sensitive master by 61\% against RR and 45\% against
QoS and lowers single-beat bulk latency by 83\% at full load. Whether that cost is worth paying depends on the traffic. It is returned
only above the saturation knee and when a latency-critical master
shares the slave with bulk bursts. Below the knee and without that mix,
the three behave alike and the cheaper one is the right
choice.

\section{Conclusion}\label{sec:conclusion}

We implemented three service architectures for a 3×3 AXI4 crossbar in ASAP7 and measured both their transaction behavior under contention and their physical cost. Round-robin arbitrates per request, so a burst-heavy master takes bandwidth in proportion to its burst length and a latency-critical master waits its turn regardless of priority. QoS with aging serves the priority master first but cannot stop a long burst already in service, and leaves bandwidth share untouched. QoS-aware DRR charges each grant in beats, so a latency-critical master meets no long-burst rival, and it brings two equal-weight bulk masters to beat parity at full contention. It reaches that parity without modifying the transactions so each burst is admitted or held as a whole and reaches the slave as issued. Priority and bandwidth share stay separate, one carried per transaction on the bus and one held per master in configuration, so a master can be given low latency without a large share, or a large share without priority. 

QoS-aware DRR has the longest critical path and the most wire length of the three, yet it delivers the lowest service latency to the latency-critical master at the same clock, because the reduction in cycles a request waits outweighs the added delay per cycle. The added area and power buy a scheduling behavior that the faster, smaller architectures cannot reach at any clock. RR is the cheapest and suffices when contention is light or masters are symmetric and QoS with aging buys latency ordering at moderate cost. In an interconnect where a real-time client shares a slave with DMA or accelerator traffic, whose contention peaks above the saturation knee, QoS-aware DRR is the one of the three that serves both classes of traffic in that regime, and the measured cost figures give the price of choosing it.

\bibliographystyle{IEEEtran}
\bibliography{references}

@manual{arm_axi,
  author       = {{Arm Limited}},
  title        = {{AMBA} {AXI} and {ACE} Protocol Specification: {AXI3}, {AXI4}, and {AXI4-Lite}, {ACE} and {ACE-Lite}},
  organization = {Arm Limited},
  number       = {ARM IHI 0022E},
  year         = {2013},
  note         = {Issue E}
}

@manual{arm_qos400,
  author       = {{Arm Limited}},
  title        = {{CoreLink} {QoS-400} Network Interconnect Advanced Quality of Service, Supplement to {CoreLink} {NIC-400} Network Interconnect Technical Reference Manual},
  organization = {Arm Limited},
  number       = {ARM DSU 0026F},
  year         = {2016},
  note         = {Revision r1p0}
}

@article{restuccia2019fair,
  author  = {Restuccia, Francesco and Pagani, Marco and Biondi, Alessandro and Marinoni, Mauro and Buttazzo, Giorgio},
  title   = {Is Your Bus Arbiter Really Fair? {R}estoring Fairness in {AXI} Interconnects for {FPGA} {SoCs}},
  journal = {ACM Transactions on Embedded Computing Systems},
  volume  = {18},
  number  = {5s},
  pages   = {51:1--51:22},
  year    = {2019},
  doi     = {10.1145/3358183}
}

@inproceedings{pagani2019,
  author    = {Pagani, Marco and Rossi, Enrico and Biondi, Alessandro and Marinoni, Mauro and Lipari, Giuseppe and Buttazzo, Giorgio},
  title     = {A Bandwidth Reservation Mechanism for {AXI}-Based Hardware Accelerators on {FPGAs}},
  booktitle = {31st Euromicro Conference on Real-Time Systems (ECRTS 2019)},
  series    = {Leibniz International Proceedings in Informatics (LIPIcs)},
  volume    = {133},
  pages     = {24:1--24:24},
  year      = {2019},
  doi       = {10.4230/LIPIcs.ECRTS.2019.24}
}

@inproceedings{slijepcevic2017,
  author    = {Slijepcevic, Mladen and Hernandez, Carles and Abella, Jaume and Cazorla, Francisco J.},
  title     = {Design and Implementation of a Fair Credit-Based Bandwidth Sharing Scheme for Buses},
  booktitle = {Design, Automation \& Test in Europe Conference \& Exhibition (DATE)},
  pages     = {926--929},
  year      = {2017},
  doi       = {10.23919/DATE.2017.7927122}
}

@inproceedings{benz2024realm,
  author    = {Benz, Thomas and Ottaviano, Alessandro and Balas, Robert and Garofalo, Angelo and Restuccia, Francesco and Biondi, Alessandro and Benini, Luca},
  title     = {{AXI-REALM}: A Lightweight and Modular Interconnect Extension for Traffic Regulation and Monitoring of Heterogeneous Real-Time {SoCs}},
  booktitle = {Design, Automation \& Test in Europe Conference \& Exhibition (DATE)},
  pages     = {1--6},
  year      = {2024},
  doi       = {10.23919/DATE58400.2024.10546511}
}

@article{benz2025realm,
  author  = {Benz, Thomas and Ottaviano, Alessandro and Liang, Chaoqun and Balas, Robert and Garofalo, Angelo and Restuccia, Francesco and Biondi, Alessandro and Rossi, Davide and Benini, Luca},
  title   = {{AXI-REALM}: Safe, Modular and Lightweight Traffic Monitoring and Regulation for Heterogeneous Mixed-Criticality Systems},
  journal = {IEEE Transactions on Computers},
  volume  = {74},
  number  = {9},
  pages   = {3072--3086},
  year    = {2025},
  doi     = {10.1109/TC.2025.3584530}
}

@article{jiang2023axiicrt,
  author  = {Jiang, Zhe and Yang, Kecheng and Fisher, Nathan and Gray, Ian and Audsley, Neil C. and Dong, Zheng},
  title   = {{AXI-IC$^{\mathrm{RT}}$}: Towards a Real-Time {AXI}-Interconnect for Highly Integrated {SoCs}},
  journal = {IEEE Transactions on Computers},
  volume  = {72},
  number  = {3},
  pages   = {786--799},
  year    = {2023},
  doi     = {10.1109/TC.2022.3179227}
}

@article{kurth2022,
  author  = {Kurth, Andreas and R{\"o}nninger, Wolfgang and Benz, Thomas and Cavalcante, Matheus and Schuiki, Fabian and Zaruba, Florian and Benini, Luca},
  title   = {An Open-Source Platform for High-Performance Non-Coherent On-Chip Communication},
  journal = {IEEE Transactions on Computers},
  volume  = {71},
  number  = {8},
  pages   = {1794--1809},
  year    = {2022},
  doi     = {10.1109/TC.2021.3107726}
}

@article{shreedhar1996,
  author  = {Shreedhar, M. and Varghese, George},
  title   = {Efficient Fair Queuing Using Deficit Round-Robin},
  journal = {IEEE/ACM Transactions on Networking},
  volume  = {4},
  number  = {3},
  pages   = {375--385},
  year    = {1996},
  doi     = {10.1109/90.502236}
}

@inproceedings{demers1989,
  author    = {Demers, Alan and Keshav, Srinivasan and Shenker, Scott},
  title     = {Analysis and Simulation of a Fair Queueing Algorithm},
  booktitle = {Proceedings of the ACM SIGCOMM Symposium on Communications Architectures \& Protocols},
  pages     = {1--12},
  year      = {1989},
  doi       = {10.1145/75246.75248}
}

@book{silberschatz2018,
  author    = {Silberschatz, Abraham and Galvin, Peter B. and Gagne, Greg},
  title     = {Operating System Concepts},
  edition   = {10th},
  publisher = {Wiley},
  address   = {Hoboken, NJ, USA},
  year      = {2018}
}

@article{clark2016asap7,
  author  = {Clark, Lawrence T. and others},
  title   = {{ASAP7}: A 7-nm {finFET} Predictive Process Design Kit},
  journal = {Microelectronics Journal},
  volume  = {53},
  pages   = {105--115},
  year    = {2016},
  doi     = {10.1016/j.mejo.2016.04.006}
}

@article{bresenham1965,
  author  = {Bresenham, Jack E.},
  title   = {Algorithm for Computer Control of a Digital Plotter},
  journal = {IBM Systems Journal},
  volume  = {4},
  number  = {1},
  pages   = {25--30},
  year    = {1965},
  doi     = {10.1147/sj.41.0025}
}

@misc{st_stm32h743,
  author       = {{STMicroelectronics}},
  title        = {{STM32H743/753} {Arm} {Cortex-M7} Microcontrollers}, howpublished = {\url{https://www.st.com/en/microcontrollers-microprocessors/stm32h743-753.html}}
}

\end{document}